\documentstyle[12pt]{article}
\begin {document}
\setcounter{page}{0}

\title{Mass spectra and leptonic decay widths of heavy
quarkonia\footnote{Partly supported by the KBN grant 2P30207607}}

 \author{L. Motyka and K. Zalewski\thanks{Also at the Institute of Nuclear
 Physics, Krak\'ow, Poland}\\
 Institute of Physics, Jagellonian University, Krak\'ow, Poland}

 \maketitle

\begin{abstract}
{A nonrelativistic Hamiltonian with plausible spin dependent corrections is
proposed for the quarkonia below their respective strong decay thresholds. With
only six free parameters this model reproduces the nine known masses of the
bottomonia within about 1 MeV, the six known masses of the charmonia within a
few MeV and the five known leptonic decay widths of the ${}^3S_1$ states within
about \mbox{20 \%}. The model is then used to predict the masses of the
remaining 43 quarkonia (some of them for the first time) and of the leptonic
decay widths of the two ${}^1S_0(\overline{b}c)$ states. Comparison with some
other models is made.}
\end{abstract}

\newpage

\section{Introduction}

In the present paper we study the mass spectra of heavy
quarkonia below their strong decay thresholds, i.e. below 10558 MeV for the
$\overline{b}b$ quarkonia, below 7144 MeV for the $\overline{b}c$ quarkonia and
below 3729 MeV for the $\overline{c}c$ quarkonia. For the quarkonia with
unnatural parity, which because of the conservation laws cannot decay
strongly into two pseudoscalar mesons, these thresholds should be a little
higher. In practice this distinction is important only for the quarkonia
$2P_1(\overline{b}c)$ and $2P_{1'}(\overline{b}c)$, which according to our
calculation have masses above the minimal thresholds given above, but below
their real threshold, which is 7189 MeV. As a byproduct we obtain
the leptonic widths of the ${}^3S_1$ states of the $\overline{b}b$
and of the $\overline{c}c$ quarkonia, as well as of the ${}^1S_0$ states of the
$\overline{b}c$ quarkonia. Experimentally, out of the 34 $\overline{b}b$
quarkonia expected nine have been observed. Here and in the following we
consider a particle as observed, if it is listed as firmly established in the
1996 Particle Data Group Tables \cite{PDG}. Out of the eight expected
$\overline{c}c$ quarkonia six have been observed and only the singlet $P$ state
and the excited $\eta_c$ are still missing. The masses of both have been
reported \cite{PDG}, but they are not considered as firmly established. None of
the $\overline{b}c$ quarkonia has been observed, but candidates have been
reported \cite{BRU} and discoveries are expected in the near future.

In our previous paper \cite{MOZ} (further quoted I) we have pointed out that a
simple nonrelativistic model can reproduce, among other things, the masses of
the known ${}^3S(\overline{b}b)$ states and of the centres of gravity of the
known ${}^3P(\overline{b}b)$ states within the experimental errors. After this
paper had been published, we became aware of a series of papers (cf. ref.
\cite{CCO} and papers quoted there) using essentially the same potential. There
are many models, which give good fits to the masses of the $\overline{b}b$
quarkonia. The review paper \cite{BES} quotes and discusses about 30 of them.
Our model, however, seems to be the only one so far, which reproduces these
masses within the experimental errors, i.e. with a precision of about 0.5 MeV,
which corresponds to 50 ppm. of the total mass, or to 0.1 per cent of the first
excitation energy. This result is amazing, because the mean square velocity of
a $b$-quark in the ground state of the $\overline{b}b$ system is, in a system
of units where the velocity of light $c=1$, $\langle v^2 \rangle \approx 0.08$.
Thus a simple-minded estimate of the relativistic corrections would give
$\langle v^2 \rangle^2 \approx 0.6\%$ of the total mass. The usual
interpretation is that the nonrelativistic Hamiltonian is an effective
Hamiltonian and that the relativistic corrections are taken into account by a
renormalization of the parameters of this Hamiltonian.

In the present paper we extend the model described in I in two ways. Firstly,
we generalize the nonrelativistic potential so that it applies also to
$\overline{b}c$ and $\overline{c}c$ quarkonia. This requires one more parameter
--- the mass of the $c$-quark. We find that the predictions for the
$\overline{c}c$ quarkonia agree with experiment within about 4 MeV i.e. within
about 0.1 per cent of the total mass, or one per cent of the first
excitation energy. Since the mean square velocity of the $c$-quark in the
ground state of the $\overline{c}c$ system is about 0.25, i.e. the mean root
square velocity is about half the velocity of light, this good fit is even more
striking than that for the $\overline{b}b$ case. Then we make predictions for
the yet undiscovered states hoping that, since our fits for the known states
are good, the predictions will also work. One should keep in mind, however,
that the problem of the $\overline{b}c$ quarkonia is not just an extrapolation
between the $\overline{b}b$ and $\overline{c}c$ cases. If our handling of the
effects of the mass difference between the two constituents is faulty, the error
of the predictions may be larger than expected.

Secondly, we supplement our nonrelativistic Hamiltonian with the standard
spin-dependent terms. With one more parameter -- the coupling constant
$\alpha_s(m_c^2)$ -- we describe all the hyperfine and fine splittings, as well
as the leptonic decay widths. The value of $\alpha_s(m_c^2)$ is so plausible --
it corresponds to $\alpha_s(m_Z^2) = 0.115$ -- that one can either interpret it
as a determination of $\alpha_s(m_Z^2)$, or as a known quantity and not a free
parameter. We choose the former possibility. By comparison with experiment
we find that the errors of the calculated splittings do not exceed about 1 MeV
for the
$\overline{b}b$
systems and about 5 MeV for the $\overline{c}c$ systems. Also the leptonic
decay widths agree reasonably well with experiment. Again by analogy we expect
that our predictions for the yet undiscovered states should be good.

\section{Spin averaged masses of spin triplet states}

The nonrelativistic potential used in the present paper for the colour-singlet
$Q\overline{Q}$-systems, where $\overline{Q}$ may, but does not have to, be the
charge conjugate of $Q$, is

\begin{equation}
\label{potnr}
V(r) = m_{\overline{Q}} + m_Q - 0.78891 + 0.70638\sqrt{r} -
0.32525\frac{1}{r},
\end{equation}
where all the constants are in suitable powers of GeV and the quark masses are

\begin{equation}
\label{qumass}
m_b = 4.8030\mbox{ GeV},  \hspace{1cm} m_c = 1.3959\mbox{ GeV}
\end{equation}
For the $\overline{b}c$ quarkonia we use the reduced mass

\begin{equation}
\mu_{bc} = \frac{m_b m_c}{m_b + m_c} = 1.0816\mbox{ GeV}.
\end{equation}
The parameters are given with so many digits only in order to assist the
reader, who would like to check our calculations. We do not claim to have
established the quark masses with such accuracies. Similar remarks apply to
coupling constants etc. given further. For the $\overline{b}b$ system this
potential reduces to the potential given in I.

According to our interpretation the eigenvalues of the nonrelativistic
Hamiltonian with this potential should be interpreted as the masses of the
centres of gravity of the spin triplets. Another popular interpretation is that
the nonrelativistic Hamiltonian should give the centres of gravity of the full
spin multiplets. Let us consider first the $\overline{b}b$ and $\overline{c}c$
quarkonia. For the $L>0$ states in our model and in many others the mass of the
spin singlet component of each multiplet coincides with the centre of gravity
of the spin triplet. Thus the two interpretations are equivalent. The
difference occurs only for the $S$-states. Since the success of the
nonrelativistic models depends on a cancellation of errors, which is not
understood, we cannot give a rigorous argument to justify one interpretation
rather than the other. Let us quote, however, two plausibility arguments.
Roncaglia and collaborators \cite{RDL} explain that the spectrum of triplet
states should be particularly regular. We chose the centres of gravity of the
triplets for the practical reasons that the masses of the spin singlets for the
$\overline{b}b$ quarkonia are not known. {\it A posteriori}, however, we see
that this choice has worked, which supports the conjecture that it is an
acceptable one. For the $\overline{b}c$ quarkonia the situation is even more
confused, because spin, in general, is not a good quantum number. We assume
tentatively that the eigenvalues of the nonrelativistic Hamiltonian  correspond
to the centres of gravity of the spin triplets, when the singlet-triplet mixing
is switched off. For $L>0$ the difference between these centres of gravity and
the centres of gravity of the whole spin multiplets are small -- 3 MeV, 2 MeV
and 0.8 MeV for the $1P$, $2P$ and $1D$ multiplets respectively. For the
$S$-states the situation should be similar to that for the other quarkonia.

The results of our calculation for the ten $\overline{b}b$ masses of interest
are given in Table 1. In the five cases, where comparison with experiment is
possible, agreement is very good. A comparison with some models is also shown
in the Table. Eichten and Quigg \cite{EIQ} use the potential of Buchm\"uller
and Tye \cite{BUT}. This potential is particularly respectable, because at
short distances it reproduces correctly the two-loop result from QCD. Their
agreement with experiment is very good for the low masses, but for the higher
masses it deteriorates, with the error reaching almost 30 MeV for the $2P$
state. Kwong and Rosner \cite{KWR} use the masses of the known $S$ and $P$
states as input, calculate from this input the potential and from this
potential the masses of the other states. Thus their method avoids the bias due
to a preconceived form of the potential. Their results for the centres of
gravity of the triplets agree in general with ours, the greatest difference
being 6 MeV for the $3P$ states.

The results for the $\overline{c}c$ quarkonia are shown in Table 2. It is seen
that the approach of Eichten and Quigg is doing much better here, about as well
as ours. The results of Gupta and Johnson are the best.

The results for the $\overline{b}c$ quarkonia are given in Table 3. Since there
are no experimental data, we included more theoretical predictions. Many others
can be found from the references given in the papers used in our tables. Of
particular interest is the comparison of our results with those of Gupta and
Johnson \cite{GUJ}. Their fit is only slightly worse than ours for the
$\overline{b}b$ quarkonia and slightly better than ours for the $\overline{c}c$
quarkonia. On the whole the quality of the two fits is good (errors not
exceeding a few MeV) and comparable. Nevertheless, the physical assumptions
behind them are very different. In particular Gupta and Johnson use the
complete relativistic expression for the kinetic energy of the quarks. They
also use many more free parameters than we do. For the $\overline{b}c$
quarkonia, as shown in Table 3, the predictions of the two approaches differ
significantly. The masses of the ${}^3S$ states calculated by Gupta and Johnson
are larger than ours by about 40 MeV. For the $1P$ state the difference is in
the same direction, but smaller -- about 16 MeV. The potential in the model of
Yu-Qi Chen and Yu-Ping Kuang \cite{CHK} is a modification of the potential of
Buchm\"uller and Tye. We quote only one of the several versions, which they
propose. The model of Roncaglia et al. \cite{RDL} does not use a potential, but
obtains the masses by assuming that for a given kind of triplet resonances
(e.g. for ground states) the mass of the particle is a simple function of the
reduced mass of its constituent quarks. Then the masses of the unknown
particles are obtained by interpolation or extrapolation from the known masses.
Gershtein et al. \cite{GKL} use the Martin potential \cite{MAR} supplemented
with some relativistic and QCD inspired corrections. The scatter of the
predictions is of some tens of MeV and the two models, which agree particularly
well with the data for the $\overline{b}b$ and $\overline{c}c$ quarkonia, i.e.
that of Gupta and Johnson and ours, are not close to each other in their
predictions here. A comparison with the experimental data, when they come, will
be, therefore, of great interest.

\section{Hyperfine splittings}

The spin dependent correction to the nonrelativistic Hamiltonian, which is
responsible for the hyperfine splitting of the mass levels, is generally used
in the form (cf. e.g. \cite{BUH})

\begin{equation}
\label{HHF}
H_{HF} = \frac{32\pi\alpha_s}{9 m_Q m_{\overline{Q}}}({\bf s_1}\cdot{\bf s_2} -
\frac{1}{4}) \delta({\bf r}),
\end{equation}
adapted from the Breit-Fermi Hamiltonian. The number $\frac{1}{4}$ subtracted
from the product of the spins corresponds to our assumption that the
unperturbed nonrelativistic Hamiltonian gives the energy of the triplet. Since
for the states with orbital angular momentum $L > 0$ the wave function vanishes
at the origin, the shift affects only the $S$ states. Thus, the only first
order effect of the perturbation (\ref{HHF}) is to shift the ${}^1S_0$ states
down in energy by

\begin{equation}
\Delta E_{HF} = \frac{32\pi\alpha_s}{9 m_Q m_{\overline{Q}}}|\psi({\bf 0})|^2.
\end{equation}
In order to apply this formula one needs the value of the wave function at the
origin -- this is obtained by solving the Schr\"odinger equation with the
nonrelativistic Hamiltonian -- and the coupling constant $\alpha_s$.

Like most authors (cf. e.g. \cite{EIQ}), we determine the coupling constant
$\alpha_s(m_c^2)$ from the well measured hyperfine splitting of the $1
S(\overline{c}c)$ state. The experimental value \cite{PDG} $117\pm 2$ MeV
yields

\begin{equation}
\label{asmc2}
\alpha_s(m_c^2) = 0.3376.
\end{equation}
Actually, the experimental uncertainty of the measured hyperfine splitting
introduces an uncertainty in this value, but we think that other uncertainties
in our calculation are more serious and we do not keep track of this particular
uncertainty. Knowing the coupling at the scale $m_c^2$ we obtain the couplings
at other scales as follows. The formula including the NNLO terms from
\cite{PDG} is used to correlate $\alpha_s(\mu^2)$ with the parameters
$\Lambda^{(n_f)}$. The number of flavours $(n_f)$ is put equal to three for
$\mu^2 \leq m_c^2$ (we are not interested in the region $\mu^2 \leq m_s^2$),
equal to four for $m_b^2 \geq \mu^2 \geq m_c^2$ and equal to five for $\mu^2
\geq m_b^2$ (we are not interested in the region $\mu^2 \geq m_t^2$). Then the
value of $\alpha_s(m_c^2)$ from (\ref{asmc2}) is used to calculate
$\Lambda^{(3)}$ and $\Lambda^{(4)}$. Using the known value of $\Lambda^{(4)}$
and the formula from ref \cite{PDG} we find the value

\begin{equation}
\label{asmb2}
\alpha_s(m_b^2) = 0.2064
\end{equation}
From this the value of $\Lambda^{(5)}$ is found and further
$\alpha_s(m_Z^2)$ is calculated. The value $\alpha_s(m_Z^2) = 0.115$ obtained
from this calculation agrees very well with the other determinations of this
parameter compiled by the Particle Data Group \cite{PDG}. Note that this
supports our model, since a different choice of the Hamiltonian would in
general lead to a different value of the wave function at the origin and to a
different determination of $\alpha_s(m_c^2)$ from the same hyperfine splitting.
Then the estimate of $\alpha_s(m_Z^2)$ would, of course, be also different.
For the hyperfine splitting of the $\overline{b}c$ quarkonia we use the
coupling constant

\begin{equation}
\alpha_s(4\mu_{bc}^2) = 0.2742,
\end{equation}
so that in each case the scale is twice the reduced mass of the quark-antiquark
system.

The calculated hyperfine splittings are given in Tables 1--3. No confirmed
experimental data to check these predictions are available as yet. Let us note,
however, that the unconfirmed experimental splitting of the $2S(\overline{c}c)$
level -- 92 MeV -- is much bigger than expected from the models. In all cases,
where comparison with the other models is possible, the hyperfine splittings
predicted from our model are significantly smaller than the splittings found by
Eichten and Quigg \cite{EIQ} and similar to, but usually a little larger than,
the splittings calculated by Gupta and Johnson \cite{GUJ}.

One can also try to compare our results with more ambitious approaches. A
careful analysis in the framework of QCD sum rules \cite{NAR} finds for the
hyperfine splitting of the $1S(\overline{b}b)$ state $63^{+29}_{-51}$ MeV. The
central value agrees very well with our expectation, but the uncertainty is too
large to distinguish between the potential models. A lattice calculation
\cite{DAV} gives for the hyperfine splitting of the $1S(\overline{b}c)$ state
$60$ MeV with a large uncertainty. Again the central value is close to our
model, but the uncertainty is big enough to be consistent with all the
potential models quoted here.

Let us conclude this section with two comments. The operator $H_{HF}$, besides
shifting the ${}^1S_0$ energy levels by its diagonal matrix elements, mixes
the ${}^1S_0$ states corresponding to various principal quantum numbers.
Formally this gives significant corrections to the energy levels. These
corrections, however, are second and higher order in the perturbation. We
follow the usage of assuming that they cancel with other second and higher
order corrections beyond our control. At first order there are only small
admixtures of other ${}^1S_0$ states in any given ${}^1S_0$ state. These,
however, seem of little interest. The QCD corrections to hyperfine splittings
have been calculated in various approximations (cf. ref. \cite{CHO} and
references given there). Since these corrections are small and controversial
(it has been argued that they cancel with other correction \cite{CHO}), we do
not include them in our model.

\section{Leptonic decay widths}

The leading terms in the leptonic decay widths of the heavy quarkonia are
proportional to the squares of the wave functions at the origin. Therefore,
they are significant only for the $S$ states. For the $\overline{b}b$ quarkonia
and the $\overline{c}c$ quarkonia we shall consider the decays of the ${}^3S$
(vector) states into pairs of charge conjugated charged leptons, e.g. for
definiteness into $e^+e^-$ pairs. For the $\overline{b}c$ quarkonia we consider
the decays of the ${}^1S$ (pseudoscalar) states into $\tau\nu_\tau$ pairs.
Since the probability of such decays contains as a factor the square of the
lepton mass, the decays into lighter leptons are much less probable.

The decay widths of the vector $\overline{b}b$ and $\overline{c}c$ quarkonia
into charged lepton pairs are usually calculated from the QCD corrected Van
Royen - Weisskopf formula \cite{VRW}, \cite{BAR}

\begin{equation}
\Gamma_{V\rightarrow \overline{l}l} = 16\pi\alpha^2e_Q^2 \frac{|\psi({\bf
0})|^2}{M_V^2}\left( 1 - \frac{16\alpha_s(m_Q^2)}{3\pi}\right).
\end{equation}
For vector mesons containing light quarks this formula leads to paradoxes (cf.
\cite{NAZ} and references contained there). For quarkonia, however, the main
problem seems to be the QCD correction. Using the coupling constants
$\alpha_s(m_Q^2)$ found in the preceding section one finds that the correction
linear in $\alpha_s$ is 57 per cent for $\overline{c}c$ and 35 per cent for
$\overline{b}b$. Thus in order to get quantitative predictions it is necessary
to include higher order corrections, which, however, are not known.
In order to guestimate the missing terms we tried two simple Ans\"atze.
Exponentialization of the first correction yields

\begin{equation}
C_1(\alpha_s(m_Q^2)) = \exp\left( -\frac{16\alpha_s(m_Q^2)}{3\pi}\right),
\end{equation}
while Pad\'eization gives
\begin{equation}
C_2(\alpha_s(m_Q^2)) = \frac{1}{1 + \frac{16\alpha_s(m_Q^2)}{3\pi}}.
\end{equation}
We use the arithmetic average of these two estimates as our estimate of the QCD
correction factor extended to higher orders. The difference between $C_1$
and $C_2$ is our crude evaluation of the uncertainty of this estimate. The
resulting leptonic widths are collected in Table 4. Combining in quadrature the
experimental errors with our estimates of the theoretical uncertainties we get
a good overall agreement ($\chi^2/ND = 5.9/5$). About half of the $\chi^2$,
however, comes from the decay width of the $2^3S(\overline{b}b)$, where the
predicted value is significantly larger than the newly included experimental
value \cite{PDG}. Thus here there may be a problem. Let us note the relation

\begin{equation}
\Gamma_{V\rightarrow \overline{l}l} = \frac{9}{8}\; \frac{4 m_Q^2}{M_V^2}
\frac{\alpha^2 e_Q^2}{\alpha_s(m_Q^2)} C_{av} \Delta E_{HF},
\end{equation}
where $C_{av}$ is the QCD correction factor. With our choice of parameters this
formula reduces to

\begin{equation}
\Gamma_{V\rightarrow \overline{l}l} = F(Q) \frac{4 m_Q^2}{M_V^2} \Delta E_{HF},
\end{equation}
with $F(c) = 4.73\cdot 10^{-5}$ and $F(b)= 2.33\cdot 10^{-5}$.

The formula for the leptonic widths of the pseudoscalar $\overline{b}c$
quarkonia reads

\begin{equation}
\Gamma_{\tau\nu} = \frac{G^2}{8\pi} f_{B_c}^2 |V_{cb}|^2 M_{B_c} m_\tau^2
\left( 1 - \frac{m_\tau^2}{M_{B_c}^2}\right)^2,
\end{equation}
where $G$ is the Fermi constant, $V_{cb} \approx 0.04$ is the element of the
Cabibbo-Kobayashi-Masakawa matrix and the decay constant $f_{B_c}$ is given by
the formula (cf. e.g. \cite{BRF})

\begin{equation}
f_{B_c}^2 = \frac{12 |\psi({\bf 0})|^2}{M_{B_c}}\overline{C}^2(\alpha_s),
\end{equation}
where $\overline{C}(\alpha_s)$ is a QCD correction factor. Formally this decay
constant is defined in terms of the element of the axial weak current

\begin{equation}
\langle0|A_\mu(0)|B_c(q)\rangle = i f_{B_c} V_{cb} q_\mu.
\end{equation}
Thus it corresponds to $f_\pi \approx 131$ MeV. The QCD correction factor is
\cite{BRF}
\begin{equation}
\overline{C}(\alpha_s) = 1 -
\frac{\alpha(4\mu_{bc}^2)}{\pi} \left[ 2 -
\frac{m_b - m_c}{m_b + m_c} \log\frac{m_b}{m_c}\right].
\end{equation}
With our parameters $\overline{C}(\alpha_s) \approx 0.885$ and since this is
rather close to unity, we use it without trying to estimate the higher order
terms.

Substituting the numbers one finds the decay widths given in Table 4. The
corresponding decay constants for the ground state and for the first excited
$S$-state of the $\overline{b}c$ quarkonium are $f_{B_c} = 435 MeV$ and
$f_{B_c} = 315$ MeV.

Let us note the convenient relation

\begin{equation}
f_{B_c}^2 = \frac{27 \mu_{bc}}{8\pi\alpha_s(4\mu_{bc}^2)} \frac{m_b +
m_c}{M_{B_c}} \overline{C}^2(\alpha_s)\Delta E_{HF},
\end{equation}
which for our values of the parameters yields

\begin{equation}
f_{B_c} = 57.6 \sqrt{\frac{6199}{M_{B_c}}} \sqrt{\Delta E_{HF}},
\end{equation}
where all the parameters are in suitable powers of MeV.

\section{Fine structure of the levels}

The spin dependent correction to the nonrelativistic Hamiltonian, which is
responsible for the fine splittings, is also modelled on the Breit-Fermi
Hamiltonian \cite{EIQ}, \cite{GKL}. It can be decomposed into a part, which is
antisymmetric with respect to the spins of the constituents

\begin{equation}
V_A(r) = \frac{1}{4}\left( \frac{1}{m_Q^2} - \frac{1}{m_{\overline{Q}}^2}
\right)\left(-\frac{1}{r}\frac{dV(r)}{dr} + \frac{8 \alpha_s}{3 r^3}\right)
{\bf L}\cdot({\bf s}_Q  - {\bf s}_{\overline{Q}}),
\end{equation}
where $V(r)$ is the nonrelativistic potential (\ref{potnr}), and a part
symmetric in these spins. The symmetric part can be decomposed into a
spin-orbit coupling

\begin{equation}
V_{LS}(r) = \frac{1}{4}\left(\frac{1}{m_Q^2} + \frac{1}{m_{\overline{Q}}^2}
\right)\left(-\frac{1}{r}\frac{dV(r)}{dr} + \frac{8 \alpha_s}{3 r^3}\right)
{\bf L\cdot{}S} + \frac{4\alpha_s}{3 m_Q m_{\overline{Q}}}\frac{1}{r^3}{\bf
L\cdot{}S},
\end{equation}
where

\begin{equation}
{\bf S} = {\bf s}_Q  + {\bf s}_{\overline{Q}},
\end{equation}
and a tensor part

\begin{equation}
V_T(r) = \frac{4 \alpha_s}{3} \frac{1}{r^3} [3({\bf s}_Q\cdot\hat{\bf n})
({\bf s}_{\overline{Q}}\cdot\hat{\bf n}) -{\bf s}_Q\cdot{\bf
s}_{\overline{Q}}],
\end{equation}
where the versor $\hat{\bf n} = \frac{{\bf r}}{r}$. In the first perturbative
approximation these corrections to the Hamiltonian not only shift the mass
levels, but also mix some states with different values of the orbital angular
momentum and spin.

Let us note some useful selection rules. Angular momentum and parity are good
quantum numbers, therefore states with different $J$ and/or $P$ do not mix.
States with orbital angular momenta differing by one unit do not mix, because
parity is a good quantum number, and states with angular momenta differing by
more than two units cannot mix because of the Eckart-Wigner theorem. From
symmetry with respect to the exchange of spins in the LS basis, the operator
$V_A$ can only contribute to matrix elements between the spin singlet and spin
triplet states. The other two operators contribute only to the matrix elements
between spin triplet states. When $\overline{Q}$ is the charge conjugate of
$Q$, charge conjugation $C = (-1)^{L+S}$ is a good quantum number and states
with different spins do not mix. The matrix elements between spin singlet
states vanish; thus, when spin is a good quantum number, we predict within each
spin multiplet with $L > 0$ :

\begin{equation}
M({}^1L) = M_{c.o.g.}({}^3L),
\end{equation}
Here the centre of gravity of the triplet is defined by the usual formula
\begin{equation}
M_{c.o.g.}({}^3L) = \frac{(2L+3)M(^3L_{L+1}) + (2L+1)M(^3L_L) +
(2L-1)M(^3L_{L-1})}{3(2L+1)}.
\end{equation}
There is no firmly established experimental data to test this prediction, but
most models and the preliminary data for the $1^1P_1(\overline{c}c)$ state
agree with it within about 1 MeV. Of course, this prediction is common to all
the models, which use the spin dependent corrections to the Hamiltonian as
given here.

Since the wave function of the quarkonium in the $LS$ basis factorizes into a
space part and a spin part, the matrix elements of the space operators and of
the spin operators can be calculated separately. The necessary matrix elements
for the spin operators between spin triplet states are

\begin{eqnarray}
\langle J,L',1|{\bf L\cdot{}S}|J,L,1\rangle =
\frac{1}{2}[J(J+1)-L(L+1)-2]\delta_{LL'}\\
\langle L+1,L,1|3({\bf s}_Q\cdot\hat{\bf n})
({\bf s}_{\overline{Q}}\cdot\hat{\bf n})
-{\bf s}_Q\cdot{\bf s}_{\overline{Q}}|L+1,L,1\rangle = \nonumber\\
-\frac{2L}{2L+3},\\
\langle L,L,1|3({\bf s}_Q\cdot\hat{\bf n})
({\bf s}_{\overline{Q}}\cdot\hat{\bf n})
-{\bf s}_Q\cdot{\bf s}_{\overline{Q}}|L,L,1\rangle = 2,\\
\langle L-1,L,1|3({\bf s}_Q\cdot\hat{\bf n})
({\bf s}_{\overline{Q}}\cdot\hat{\bf n})
-{\bf s}_Q\cdot{\bf s}_{\overline{Q}}|L-1,L,1\rangle = \nonumber\\
-\frac{2(L+1)}{2L-1},\\
\langle L+1,L-2,1|3({\bf s}_Q\cdot\hat{\bf n})
({\bf s}_{\overline{Q}}\cdot\hat{\bf n})
-{\bf s}_Q\cdot{\bf s}_{\overline{Q}}|L+1,L,1\rangle = \nonumber\\
 \frac{3\sqrt{L(L-1)}}{2(2L-1)}.
\label{dsmixi}
\end{eqnarray}
The only necessary nonzero matrix element between a spin singlet and spin
triplet state is

\begin{equation}
\langle L,L,0|{\bf L\cdot(s_Q - s_{\overline{Q}})}|L,L, 1\rangle =
\sqrt{L(L+1)}
\end{equation}
Let us consider now the two matrix elements in coordinate space. The
calculation of the matrix elements of the operator
$\frac{1}{r}\frac{dV(r)}{dr}$ is standard, but the calculation of the matrix
elements of the operator $\frac{\alpha_s}{r^3}$ requires an interpretation of
$\alpha_s$. We propose to interpret $\alpha_s$ as a function
$\tilde{\alpha}_s(r)$ defined as follows

\begin{equation}
\tilde{\alpha}_s(r) = \frac{12\pi}{33 - 2n_f} \frac{(\tilde{\Lambda}^{(n_f)}
 r)^2 - 1}{\log\left[(\tilde{\Lambda}^{(n_f)} r)^2\right]},
\end{equation}
where $n_f$ equals three for $r < \frac{1}{m_c}$, equals four for
$\frac{1}{m_c} < r < \frac{1}{m_b}$ and equals five for $r > \frac{1}{m_b}$.
The parameter $\tilde{\Lambda}^{(4)}$ is obtained from the conditions
$\tilde{\alpha}(\frac{1}{m_c}) = \alpha(m_c^2)$ and
$\tilde{\alpha}(\frac{1}{m_b}) = \alpha(m_b^2)$. Each of these conditions gives
a slightly different $\tilde{\Lambda}^{(4)}$. We use the geometric mean of the
two results. They are so close to each other that taking the arithmetic mean
instead of the geometrical one makes no difference within our precision.
Knowing $\tilde{\Lambda}^{(4)}$ we recalculate $\tilde{\alpha}_s(r)$ at $r =
\frac{1}{m_b}$ and $r = \frac{1}{m_c}$ and fix $\tilde{\Lambda}^{(3)}$ and
$\tilde{\Lambda}^{(5)}$ so that the function $\tilde{\alpha}_s(r)$ is
continuous at these points. We find

\begin{equation}
\tilde{\Lambda}^{(3)} = 0.1657\mbox{ GeV}, \hspace{0.5cm}\tilde{\Lambda}^{(4)}
= 0.1384\mbox{ GeV}, \hspace{0.5cm}\tilde{\Lambda}^{(5)} = 0.1015\mbox{ GeV}.
\end{equation}
Our form of the function $\tilde{\alpha}_s(r)$ is, of course, inspired by the
standard one-loop formula for $\alpha_s$. The numerator is introduced in order
to compensate the zero of the denominator at $\tilde{\Lambda}^{(3)}r = 1$. Its
exact form has little effect in the range of $r$ dominating the integrals.

The calculations for the $\overline{b}b$ and $\overline{c}c$ quarkonia, where
$C$ is a good quantum number, and for the $J = L \pm 1$ states of the
$\overline{b}c$ quarkonia, which must be spin triplets, involve only the energy
shifts due to the symmetric spin orbit interaction and to the tensor
interaction. The singlet and triplet $J = L$ states of the $\overline{b}c$
quarkonia mix under the influence of the antisymmetric spin-orbit interaction.
The results of the calculations are given in Tables 1-3. For the
$\overline{b}b$ quarkonia experimental data is available for the splittings of
the $1P$ and $2P$ states. Agreement between our model and this data is within
about 1 MeV. The agreement with other models is within 5 MeV, i.e. much better
than for the hyperfine splittings. A similar agreement with the model of Kwong
and Rosner holds for the $3P$ states, but for higher angular momenta the
discrepancies increase. We predict much larger splittings. In particular for
the $F$ states we predict a splitting of about 13 MeV, while Kwong and Rosner
expect a negligible splitting within about 1 MeV. Even for the $G$ states we
expect a splitting of about 10 MeV. Thus, the $L$-dependence of the fine
splittings is seen as an important observable to distinguish between models.
For the $1P(\overline{c}c)$ states our splittings agree with
experiment and with the very good predictions of Gupta and Johnson about as
well as for the centres of gravity of the triplets i.e. within about 5 MeV. The
splittings predicted by the model of Eichten and Quigg are too small by about a
factor of two. For the $\overline{b}c$ quarkonia our predictions for the $1P$
states agree with Gupta and Johnson within about 2 MeV except for $J=0$,
where our splitting is smaller by 10 MeV. For the $1D$ and $2P$ states there
are
only the predictions of Eichten and Quigg \cite{EIQ} and of Gershtein et al.
\cite{GKL} for comparison. There is rough qualitative agreement except for the
${}^3D_1$ state, where we predict a down shift by almost 20 MeV, while the
other models find only very small shifts.

The mixing between the spin singlet and the spin triplet states can be
parameterized in terms of mixing angles. We find $\sin\phi_{1P} = 0.374$,
$\sin\phi_{2P} = 0.385$ and $\sin\phi_{1D} = 0.244$. Thus the mixing within the
two $P$ multiplets is almost the same, while the mixing among the $D$ states is
somewhat smaller. Both these results contradict the results of Gershtein and
collaborators \cite{GKL}, who find that mixing increases when going from the
$1P$ to the $2P$ states and from the $2P$ to the $1D$ states. A possible reason
for this discrepancy is that these authors use for the mixing formulae, which
are different from ours. In particular their singlet-triplet mixing does not
vanish for $m_Q = m_{\overline{Q}}$. As compared with Eichten and Quigg, who
have calculated mixing only for the $P$ states, we have rough agreement for the
$2P$ states (they find $\sin\phi_{2P} = 0.290$), while they find much less
mixing for the $1P$ states ($\sin\phi_{1P} \approx 0.06$).

The mixing of spin triplet states differing by two units of orbital angular
momentum $(L-2,\; L)$ is small. In our model we find mixing angles of order
$10^{-3}$ or less. It seems of little interest, except that it enhances the
leptonic decay widths of the $L \geq 2$ states (cf. e.g. \cite{MOR}). This
enhancement, however, is difficult to calculate reliably, because a given high
$L$ state mixes with various $L-2$ states and the states above the strong
decay threshold are also important for this analysis.

\section{Conclusions}

We propose a model containing six free parameters: the three parameters in the
nonrelativistic potential (\ref{potnr}), the masses of the $c$ and $b$ quarks
(\ref{qumass}) and the strong coupling at the $m_c$ scale (\ref{asmc2}). This
model is applicable to all the heavy quarkonia below their strong decay
thresholds.

We obtain for the $\overline{b}b$ quarkonia 12 quantities (five spin averaged
masses, four independent mass differences due to fine splittings and three
leptonic decay widths) in good overall agreement with experiment. The least
successful predictions are for the fine structure shift of the $2{}^3P_1$
state, which is measured to be $-4.8 \pm 0.5$ MeV, while we find $-6.0$ MeV and
for the leptonic decay width of the $2^3S(\overline{b}b)$ state, where we find
$(0.59 \pm 0.03)$ KeV, while the newly included experimental result \cite{PDG}
is $(0.52 \pm 0.03)$ KeV. For the $\overline{c}c$ quarkonia we find 8
quantities, which can be compared with experiment (six masses and two leptonic
widths). Here in most cases the difference between the measured value and the
prediction exceeds the experimental error, but the errors in the mass
predictions do not exceed a few MeV and the errors in the leptonic decay widths
do not exceed 1.6 s.d.. On the whole, with 6 parameters we predict 20
quantities in good (for $\overline{b}b$) or fair (for $\overline{c}c$)
agreement with experiment. The only other model known to us, which has a
comparable record, is the model of Gupta and Johnson \cite{GUJ}, but this model
has many more free parameters.

We give predictions for the yet unmeasured masses of the quarkonia and for the
leptonic widths of the $\overline{b}c$ quarkonia. The predictions are listed in
Tables 1-4. Here we would like to make the following general remarks. Our model
predicts much larger fine splittings at high $L$ than the model of Kwong and
Rosner \cite{KWR}. We also find significantly heavier $\overline{b}c$
resonances than Gupta and Johnson \cite{GUJ}, which is remarkable, because the
two models give similar descriptions of the charmonia and of the bottomonia.
The discrepancy is in the spin averaged masses. The splittings of the levels,
except for the hyperfine splitting of the $1S$ level, are similar.

\newpage
{\bf Table captions}
\vspace{2em}

\noindent
Table 1. Mass spectrum of the $\bar{b}b$ quarkonia below the threshold
for strong decays (2$m_B$ = 10558~MeV) in MeV. $\Delta X$ denotes the
difference between the mass of particle $X$ and the centre of gravity of the
spin triplet part of the multiplet, where $X$ belongs.
\vspace{2em}

\noindent
Table 2. Mass spectrum of the $\bar{c}c$ quarkonia below the threshold
for strong decays (2$m_D$ = 3729~MeV) in MeV.
 $\Delta X$ denotes the
difference between the mass of particle $X$ and the centre of gravity of the
spin triplet part of the multiplet, where $X$ belongs.
\vspace{2em}

\noindent
Table 3. Mass spectrum of the $\bar{c}b$~($\bar{b}c$)~quarkonia below the
threshold for strong decays ($m_D+m_B$ = 7143~MeV, $m_{B^\ast} + m_D$ =
7189~MeV) in MeV. $\Delta X$ denotes the difference between the mass of
particle $X$ and the centre of gravity of the spin triplet part of the
multiplet, where $X$ belongs. (a) -- particle above its strong decay threshold.
\vspace{2em}

\noindent
Table 4. Leptonic widths in keV.

\newpage
\begin{figure}
\noindent
\hspace{-1.5cm}
Table 1.
\vspace{1ex}

\noindent
\hspace{-1.5cm}
\begin{tabular}{||c|c|c|c|c||c|c|c|c||}
\hline\hline
State & KR & EQ & Present & Exp. & State & KR & EQ & Present \\
      & \cite{KWR} & \cite{EIQ} & paper & & &\cite{KWR} & \cite{EIQ} & paper\\
\hline
$\Delta 1\,^1S_0   $  & --   & -87  & -56.7  & --   & $1\,^3D$ (c.o.g.)  &10156  &10127
&10155\\
\hline
$1\, ^3S_1     $  & --   & 9464 & 9460 & 9460 & $\Delta 1\, ^3D_3 $  & +4    & +3    & +7.5 \\
\hline
$1\, ^3P $ (c.o.g.) & 9903 & 9873 & 9900 & 9900 & $\Delta 1\, ^3D_2 $  & 0
& -1    & -2.2  \\
\hline
$\Delta 1\, ^3P_2  $  & --   & +13  & +13 & +13  & $\Delta 1\, ^3D_1 $  & -6    & -7    & -14 \\
\hline
$\Delta 1\, ^3P_1  $  & --   & -9   & -8.6   & -8   & $\Delta 1\, ^1D_2 $  & 0     & 0     & 0     \\
\hline
$\Delta 1\, ^3P_0  $  & --   & -39  & -39  & -40  & $1\,^3F$ (c.o.g.)  & 10348
& --
&10348\\
\hline
$\Delta 1\, ^1P_1  $  & 0    & 0    & 0      & --   & $\Delta 1\,^3F_4  $  & 0     & --    & +5.1 \\
\hline
$\Delta 2\,^1S_0   $  & --   & -44  & -28  & --   & $\Delta 1\,^3F_3  $  & +1    & --    & -0.9 \\
\hline
$2\, ^3S_1     $  & --   & 10007& 10023& 10023& $\Delta 1\,^3F_2  $  & 0     & --    & -7.9 \\
\hline
$2\, ^3P$ (c.o.g.)  & 10259& 10231& 10260& 10260& $\Delta 1\,^1F_3  $  & 0
&
--    & 0    \\
\hline
$\Delta 2\, ^3P_2  $  & --   & +11  & +9.1   & +9   & $2\,^3D$ (c.o.g.)  &
10441
& --    &10438\\
\hline
$\Delta 2\, ^3P_1  $  & --   & -7   & -6.0   & -5   & $\Delta 2\,^3D_3  $  & +3    & --    &+6.0  \\
\hline
$\Delta 2\, ^3P_0  $  & --   & -32  & -27  & -28  & $\Delta 2\,^3D_2  $  & 0     & --    &-1.8   \\
\hline
$\Delta 2\, ^1P_1  $  & 0    & 0    & 0      &      & $\Delta 2\,^3D_1  $  & -6    & --    &-11  \\
\hline
$\Delta 3\,^1S_0   $  & --   & -41  & -20  & --   & $\Delta 2\,^1D_2  $  & 0     & --    & 0    \\
\hline
$3\, ^3S_1     $  & --   & 10339&10355 &10355 & $1\,^3G$ (c.o.g.)  & --    & --    &10508\\
\hline
$3\, ^3P$ (c.o.g.)  & 10520& --   &10525 & --   & $\Delta 1\,^3G_4  $  & --
& --    &+3.8 \\
\hline
$\Delta 3\, ^3P_2  $  & +6   & --   & +7.3   & --   & $\Delta 1\,^3G_3  $  & --    & --    &-0.4 \\
\hline
$\Delta 3\, ^3P_1  $  & -4   & --   & -4.9   & --   & $\Delta 1\,^3G_2  $  & --    & --    &-5.4 \\
\hline
$\Delta 3\, ^3P_0  $  & -19  & --   & -22  & --   & $\Delta 1\,^1G_3  $  & --    & --    & 0   \\
\hline\cline{6-9}
$\Delta 3\, ^1P_1  $  &  0   & --   & 0      & --   &\multicolumn{4}{c}{} \\
\cline{1-5}\cline{1-5}
\end{tabular}
\end{figure}
\vspace{2em}

\noindent
\begin{figure}
Table 2.
\vspace{1ex}

\noindent
\begin{tabular}{|c|c|c|c|c|}
\hline
State              & EQ & GJ & Present & Exp.     \\
                   & \cite{EIQ} & \cite{GUJ} & paper&\\
\hline
$\Delta 1\,^1S_0    $  & -117 & -117 & -117 & -117  \\
\hline
$1\, ^3S_1      $ & 3097 & 3097 & 3097 & 3097  \\
\hline
$1\,   P$ (c.o.g.)   & 3492 & 3526 & 3521 & 3525\\
\hline
$\Delta 1\, ^3P_2 $   & +15  & +31 &+31    & +31\\
\hline
$\Delta 1\, ^3P_1   $  & -6 & -15  & -19  & -15 \\
\hline
$\Delta 1\, ^3P_0  $   & -56  & -110 & -100 & -110\\
\hline
$\Delta 1\, ^1P_1   $  & +1   & 1 & 0 & +1??      \\
\hline
$\Delta 2\,^1S_0    $  & -78  & -68 & -72 & -92??  \\
\hline
$2\, ^3S_1      $  & 3686 & 3685 & 3690 & 3686\\
\hline
\end{tabular}
\end{figure}
\vspace{1em}

\begin{figure}
\noindent
Table 3. \vspace{1ex}

\noindent
\begin{tabular}{|c|c|c|c|c|c|c|}
\hline
State             & CK   & EQ   & Ron  & Ger  & GJ   & Present    \\
                  & \cite{CHK} & \cite{EIQ} & \cite{RDL} & \cite{GKL} &
\cite{GUJ} & paper\\
\hline
$\Delta 1\,^1S_0    $ & -45  & -73  & -65  & -64  & -41  & -58   \\
\hline
$1\, ^3S_1      $ & 6355 & 6337 & 6320 & 6317 & 6308 & 6349\\
\hline
$1\,   P$ (c.o.g.)  & 6764 & 6736 & 6753 & 6728 & 6753 & 6769\\
\hline
$\Delta 1\, ^3P_2   $ & +9   & +11  & +27  & +15  & +20  & +18 \\
\hline
$\Delta 1\, P_1     $ & 0    & 0        & 0    & +1   & +4   & +2.4  \\
\hline
$\Delta 1\, P_{1'}  $ & -4   & -6   & -13  & -11  & -15  & -15 \\
\hline
$\Delta 1\, ^3P_0   $ & -36  & -36  & -93  & -45  & -64  & -54 \\
\hline
$\Delta 2\,^1S_0    $ & -27  & -43  & --   & -35  & -33  & -33   \\
\hline
$2\, ^3S_1      $ & 6917 & 6899 & 6900 & 6902 & 6886 & 6921\\
\hline
$1\, D $ (c.o.g.)   & --   & 7009 & --   & 7009 & --   & 7040\\
\hline
$\Delta 1\, ^3D_3   $ & --   & -4   & --   & +7   & --   & +9.3  \\
\hline
$\Delta 1\, D_2     $ & --   & 0    & --   & -2   & --   & 0     \\
\hline
$\Delta 1\, D_{2'}  $ & --   & +3   & --   & -8   & --   & -2.5  \\
\hline
$\Delta 1\, ^3D_1   $ & --   & +3   & --   & -1   & --   & -18 \\
\hline
$2\,  P $ (c.o.g.)  & 7160 & 7142 & --   & 7122 & --   & 7165\\
\hline
$\Delta 2\, ^3P_2^{(a)}   $ &  +6  &  +11 & --   &  +12 & --   & +13 \\
\hline
$\Delta 2\,P_1      $ &  0   &  0   & --   &  +2  & --   & +1.9  \\
\hline
$\Delta 2\,P_{1'}   $ &  -1  & -7   & --   &  -9  & --   & -11\\
\hline
$\Delta 2\, ^3P_0   $ &  -26 & -34  & --   & -34  & --   & -39 \\
\hline
\end{tabular}
\end{figure}

\newpage

\begin{figure}
\noindent
Table 4. \vspace{1ex}

\noindent
\begin{tabular}{|c|c|c|c|}
\hline
State & EQ & Present  & Experiment\\
      &\cite{EIQ}&paper&\\
\hline
$1\, ^3S_1 \;\; (\bar{c}c)$  & 8  & 4.5 $\pm$ 0.5 & 5.3 $\pm$ 0.4  \\
$2\, ^3S_1 \;\; (\bar{c}c)$  & 3.7  & 1.9 $\pm$ 0.2 & 2.1 $\pm$ 0.2  \\
\hline
$1\, ^1S_0 \;\; (\bar{c}b)$    &
                                                 $4.0 \cdot 10^{-8}$   & $2.8
\cdot 10^{-8}$ & ---          \\
$2\, ^1S_0 \;\; (\bar{c}b)$  & ---  & $1.6 \cdot 10^{-8}$ & ---            \\
\hline
$1\, ^3S_1 \;\; (\bar{b}b)$ & 1.7  & 1.36 $\pm$ 0.07 & 1.32 $\pm$ 0.05 \\
$2\, ^3S_1 \;\; (\bar{b}b)$ & 0.8  & 0.59 $\pm$ 0.03 & 0.52 $\pm$ 0.03 \\
$3\, ^3S_1 \;\; (\bar{b}b)$ & 0.6 & 0.40 $\pm$ 0.02 & 0.48 $\pm$ 0.08 \\
\hline
\end{tabular}
\end{figure}

\end{document}